\documentclass[11pt]{article}

\usepackage{amsmath,amssymb,hyperref,color}

\newcommand{\ket}[1]{\big| #1 \big\rangle}  % | >
\newcommand{\bra}[1]{\big\langle #1 \big|}  % < |
\newcommand{\braket}[2]{\big\langle #1 \big| #2 \big\rangle}                 % < | >
\newcommand{\arxiv}[1]{\emph{arXiv:\linebreak[0]\href{http://arxiv.org/abs/#1}{#1}}}

\usepackage{authblk}

\begin{document}

\title{Spatial Search on Johnson Graphs by \\ Discrete-Time Quantum Walk%\thanks{Grants or other notes
}


\author[1]{Hajime Tanaka}
\author[1]{Mohamed Sabri}
\author[2]{Renato Portugal}

%\authorrunning{Hajime Tanaka \and Mohamed Sabri \and  Renato Portugal} % if too long for running head

\affil[1]{Research Center for Pure and Applied Mathematics

          Graduate School of Information Sciences
          
          Tohoku University,          
          Sendai 980-8579, Japan 
          %\email{htanaka@tohoku.ac.jp, sabrimath@dc.tohoku.ac.jp}           %  \\
              }

\affil[2]{National Laboratory of Scientific Computing (LNCC)

          Petr\'opolis, RJ, 25651-075, Brazil
          %\\          {portugal@lncc.br}
}

%%%% The correct dates will be entered by the editor

\maketitle

\begin{abstract}
The spatial search problem aims to find a marked vertex of a finite graph using a dynamic with two constraints: (1) The walker has no compass and (2) the walker can check whether a vertex is marked only after reaching it. This problem is a generalization of unsorted database search and has many applications to algorithms. Classical algorithms that solve the spatial search problem are based on random walks and the computational complexity is determined by the hitting time. On the other hand, quantum algorithms are based on quantum walks and the computational complexity is determined not only by the number of steps to reach a marked vertex, but also by the success probability, since we need to perform a measurement at the end of the algorithm to determine the walker's position. In this work, we address the spatial search problem on Johnson graphs using the coined quantum walk model. Since Johnson graphs are vertex- and distance-transitive, we have found an invariant subspace of the Hilbert space, which aids in the calculation of the computational complexity. We have shown that, for every fixed diameter, the asymptotic success probability is $1/2$ after taking $\pi\sqrt N/(2\sqrt 2)$ steps, where $N$ is the number of vertices of the Johnson graph.

\end{abstract}

%%%%%%%%%%%%%%%%%%%
\section{Introduction}

Quantum walks (QWs) are the quantum counterpart of classical random walks. The interest in this area is increasing unremittingly for more than two decades due to its richness and because QWs are useful to build quantum algorithms and to analyze many complex quantum systems~\cite{KGK21}. The position of the walker is described by a discrete structure, such as a graph, but the time evolution can be discrete~\cite{ADZ93} or continuous~\cite{FG98}. Those versions are equivalent in some special cases~\cite{CP19}, but they seem non-equivalent in general. There are many discrete-time QW models, the most general of which are the coined~\cite{AAKV01}, the Szegedy~\cite{Sze04a}, and the staggered model~\cite{PSFG16}. They are not equivalent in their original recipe, but it is possible to extend their scope in order to establish a kind of equivalence~\cite{KPSS18}.

The coined QW model introduces an internal space for each vertex, called coin space, following the structure of classical random walks on graphs, which use a coin to specify the walker's next step or the direction of motion. The definition of the coined QW on the line was introduced in Ref.~\cite{ADZ93} and the generalization to regular graphs was presented in Ref.~\cite{AAKV01}, which have used the coin-position notation $\ket{c,v}$ for the vectors of the computational basis, where $c$ is a coin value and $v$ is a vertex. In the general case, the latter reference have used two different coin labels for one edge, keeping the formalism of undirected graphs, but using an implicit notion of a directed edge. The natural and more rigorous course of action at this point is to use the symmetric \emph{directed} graph in place of the original \emph{undirected} graph. In this case, the Hilbert space associated with the symmetric directed graph is spanned by the arc set. The walker's locations and the description of the dynamics must be changed accordingly. The arc notation and the underlying symmetric digraph was used for instance in Refs.~\cite{EHSW06,HKSS14}.

The spatial search problem was introduced by Benioff~\cite{Ben02}, who attempted to obtain a Grover-based algorithm that would find a marked location on a two-dimensional lattice faster than a random walk-based algorithm. Ambainis's algorithm for the element distinctness~\cite{Amb07a} is one of the earliest successful application of this problem using QWs instead of the Grover algorithm. Since then, it has been shown that QWs are faster than random walks on many graphs~\cite{CNAO16,AGJK20,segawa2020quantum,apers_et_al21,Marsh_2021}, but not on all of them, for instance, on a circle, QWs provide no gain comparing to random walks. The spatial search problem is simpler for vertex-transitive graphs because the computational complexity is independent of the location of the marked vertex. Johnson graphs are not only vertex-transitive but also distance-transitive, which is a property that can be used to find an invariant subspace of the Hilbert space that helps in the calculation of the computational complexity---the vertices can be partitioned into subsets of vertices that have the same distance to the marked vertex.

In this work, we present a quantum search algorithm using the standard coined QW on the Johnson graph $J(n,k)$, for arbitrary parameters $k$ and $n$. We show that for a fixed $k$ and $n\gg k$, the algorithm achieves an asymptotic success probability of $1/2$ by taking $\pi\sqrt{N}/(2\sqrt 2)$ steps, where $N$ is the number of vertices. The analysis of the computational complexity is fully rigorous from the mathematical point of view, and has borrowed some ideas from Ref.~\cite{TSP21}. The results for the discrete-time case are slightly different from the corresponding ones using the continuous-time QW because the latter case has an asymptotic success probability of $1$ by taking $\pi\sqrt{N}/2$ steps. The complexity analysis in the continuous-time case is simpler because an eigenbasis of the evolution operator can be found directly from an eigenbasis of the adjacency matrix of the Johnson graph.

The structure of this work is as follows. 
Sec.~\ref{sec: coined model} describes the evolution operator $U$ of QWs on Johnson graphs using the arc notation.
Sec.~\ref{sec: search algorithm} describes the modified evolution operator of the spatial search algorithm using the Grover coin.
Sec.~\ref{sec: invariant subspace} shows that the spatial search algorithm can be analyzed in a subspace of the Hilbert space. 
Sec.~\ref{sec: eigenbasis} finds an orthonormal eigenbasis of the invariant subspace.
Sec.~\ref{sec: eigenvectors} finds approximate eigenvectors of the modified evolution operator.
Sec.~\ref{sec: analysis} presents the analysis of the computational complexity of the search algorithm.
Sec.~\ref{sec:conc} presents our final remarks.

%%%%%%%%%%%%%%%%%%%
\section{Coined QW on $J(n,k)$ using the arc notation}\label{sec: coined model}

Let $\Gamma=(\mathcal{V},\mathcal{E})$ be the Johnson graph $J(n,k)$, where $\mathcal{V}={\mathcal{V}}(\Gamma)$ is the set of vertices and $\mathcal{E}={\mathcal{E}}(\Gamma)$ is the set of edges~\cite{BCN1989B}.
${\mathcal{V}}$ is the set of $k$-subsets of $[n]=\{1,2,\dots,n\}$, and two vertices $v,v'\in {\mathcal{V}}$ are adjacent if and only if $|v\cap v'|=k-1$. The number of vertices $N$ is $\binom{n}{k}$. Note that $J(n,k)$ is $d$-regular, where the degree $d$ is given by
\begin{equation*}
	d=k(n-k).
\end{equation*}
In this work, we assume that $n\geqslant 2k$, so that $J(n,k)$ has diameter $k$.
 Let ${\mathcal{A}}$ be the set of arcs, or directed edges, of $J(n,k)$ using the interpretation that each edge is replaced by a pair of opposite arcs. For $a\in {\mathcal{A}}$, let $\bar{a}$, $\operatorname{head}(a)$, and $\operatorname{tail}(a)$ denote the opposite arc, the head, and the tail of $a$, respectively.

The Hilbert space $\mathcal{H}_{\mathcal{A}}$ of the coined model~\cite{Portugal2018B} associated with $J(n,k)$ is spanned by the set of arcs, that is, $\mathcal{H}_{\mathcal{A}}=\operatorname{span}\!\big\{\ket{a} : a\in {\mathcal{A}}\big\}$, so that $\braket{a}{b}=\delta_{a,b}$, $\forall a,b\in{\mathcal{A}}$, and $\big\{\ket{a} : a\in {\mathcal{A}}\big\}$ is the computational basis~\cite{HKSS14}. The dimension of $\mathcal{H}_{\mathcal{A}}$ is $\left|{\mathcal{A}}\right|=2\left|{\mathcal{E}}\right|$. The evolution operator $U$ is a linear unitary operator on $\mathcal{H}_{\mathcal{A}}$ defined by 
\begin{equation*}
	U=S\,C,
\end{equation*}
where $S$ is the flip-flop shift operator, whose action on the computational basis is
\begin{equation*}
	S\ket{a}=\ket{\bar{a}} \qquad (a\in\mathcal{A}),
\end{equation*}
and $C$ is the coin operator. In this work, we use the Grover coin, widely employed in spatial search algorithms. The action of the Grover coin on the computational basis is
\begin{equation*}
	C\ket{a}= \! \sum_{\substack{b\in{\mathcal{A}}\\ \operatorname{tail}(b)=\operatorname{tail}(a)}} \!\!\!\! \left(\frac{2}{d}-\delta_{a,b}\right)\!\ket{b} \qquad (a\in\mathcal{A}).
\end{equation*}
The interpretation of the Grover coin is as follows. If the walker is on arc $a\in{\mathcal{A}}$, the action of the coin spreads the position of the walker over all arcs whose tails are equal to tail($a$), so that the amplitude of $\ket{a}$ is $(2/d-1)$ and the amplitudes of the other arcs are the same $(2/d)$. For the sake of completeness, the interpretation of the shift operator is as follows. The action of $S$ makes the walker to hop from ${a}$ to ${\bar{a}}$ and vice versa.

From the definitions of $S$ and $C$ given above, it follows that the evolution operator $U$ satisfies
\begin{equation*}
	U\ket{a} = \! \sum_{\substack{ b\in\mathcal{A} \\ \operatorname{head}(b)=\operatorname{tail}(a) }} \!\!\!\! \left( \frac{2}{d}-\delta_{\bar{a},b} \right)\! \ket{b} \qquad (a\in\mathcal{A}).
\end{equation*}

Given an initial state $\ket{\psi(0)}$, the dynamics of the model is given by 
\begin{equation*}
	\ket{\psi(t)} = U^t \ket{\psi(0)},
\end{equation*}
where $\ket{\psi(t)}$ is the state of the walk after $t$ steps ($t$ must be an integer). The probability of finding the walker on an arc $a$ after $t$ steps is
\begin{equation*}
	p_a(t)=\left|\braket{a}{\psi(t)}\right|^2,
\end{equation*}
and the probability of finding a vertex $v$ after $t$ steps is
\begin{equation*}
	p_v(t)= \!\sum_{\substack{a\in{\mathcal{A}}\\ \operatorname{tail}(a)=v}} \! \left|\braket{a}{\psi(t)}\right|^2,
\end{equation*}
which is equivalent to adding up over all coin values. If the walker is on any arc whose tail is $v$, we have enough information to find $v$. We cannot additionally consider the arcs whose heads are $v$ because that would introduce an indeterminacy.

%%%%%%%%%%%%%%%%%%%
\section{The spatial search algorithm}\label{sec: search algorithm}

Let us review the definition of the spatial search algorithm in the position-coin notation~\cite{Portugal2018B}. The Hilbert space associated with $J(n,k)$ is $\mathcal{H}_\mathcal{V}\otimes\mathcal{H}_\mathcal{C}$, where $\mathcal{H}_\mathcal{V}=\operatorname{span}\!\big\{\ket{v}:v\in \mathcal{V}\big\}$, and $\mathcal{H}_\mathcal{C}=\operatorname{span}\!\big\{\ket{c}:1\leqslant c\leqslant d\big\}$.
Let $w\in {\mathcal{V}}$ be the marked vertex. Usually, the spatial search algorithm in the coined model uses a modified coin operator $C'$ which applies the negation of the $d$-dimensional identity operator $(-I_d)$ on the marked vertex $w$ and the $d$-dimensional Grover coin $G$ on unmarked vertices, that is
\begin{equation*}
	C' = \left(I_N-\ket{w}\bra{w}\right)\otimes G+\ket{w}\bra{w}\otimes (-I_d),
\end{equation*}
where $G=2\ket{u}\bra{u}-I_d$ and\vspace{-5pt}
\begin{equation*}
	\ket{u}=\frac{1}{\sqrt{d}}\sum_{c=1}^d\ket{c}.
\end{equation*}

The dynamics of the spatial search algorithm is driven by the modified evolution operator 
$U'=SC'$.
It is straightforward to show that $U'$ can be written as 
\begin{equation*}
	U'=U R,
\end{equation*}
where $U$ is the evolution operator of the QW on $J(n,k)$ with no marked vertices and 
\begin{equation*}
	{R}=I- 2\ket{w'}\bra{w'},
\end{equation*}
where $\ket{w'}=\ket{w}\otimes\ket{u}$.
Operator $R$ is the oracle, which is able to recognize the marked vertex. 

Now let us convert the definition of the search algorithm into the arc notation. We have to redefine only $R$ because we have already defined $U$ in the arc notation. We use the same expression for ${R}=I- 2\ket{w'}\bra{w'}$, and it is enough to redefine $\ket{w'}\in \mathcal{H}_\mathcal{A}$ as
\begin{equation*}
	\ket{w'} =\frac{1}{\sqrt{d}} \! \sum_{\substack{a\in {\mathcal{A}} \\ \operatorname{tail}(a)=w}} \!\! \ket{a}.
\end{equation*}
The definition of $R$ in the arc notation is obtained by mapping $\ket{w}\otimes\ket{u}\in \mathcal{H}_\mathcal{V}\otimes\mathcal{H}_\mathcal{C}$ into $\ket{w'}\in \mathcal{H}_\mathcal{A}$.

It is standard to take the initial state of the algorithm to be the uniform superposition
\begin{equation*}
	\ket{\psi(0)} = \frac{1}{\sqrt{dN}} \sum_{a\in {\mathcal{A}}} \ket{a},
\end{equation*}
because this initial state is unbiased, so that the state of the modified QW after $t$ steps is
\begin{equation*}
	\ket{\psi(t)} = (U')^t \ket{\psi(0)}.
\end{equation*}
If the running time of the search algorithm is $t_{\mathrm{run}}$, then the success probability is given by
\begin{equation*}
	p_{\mathrm{succ}}=p_w(t_{\mathrm{run}})= \!\sum_{\substack{a\in{\mathcal{A}}\\ \operatorname{tail}(a)=w}} \! \left|\braket{a}{\psi(t_{\mathrm{run}})}\right|^2. 
\end{equation*}

For the Johnson graph $J(n,k)$, we set
\begin{equation*}
	t_{\mathrm{run}}=\left\lfloor\frac{\pi n^{k/2}}{2\sqrt{2k!}}\right\rfloor \approx \frac{\pi\sqrt{N}}{2\sqrt{2}},
\end{equation*}
where $\lfloor x\rfloor$ denotes the largest integer not exceeding $x$.
We show in this work that, for every fixed $k$, the success probability satisfies
\begin{equation*}
	p_{\mathrm{succ}}=\frac{1}{2}+ O\!\left(\frac{1}{\sqrt{n}}\right).
\end{equation*}

%%%%%%%%%%%%%%%%%%%
\section{An invariant subspace for $U'$}\label{sec: invariant subspace}

For later use, we first recall the subspace of the Hilbert space $\mathcal{H}_{\mathcal{V}}=\operatorname{span}\!\big\{\ket{v} :v\in {\mathcal{V}}\big\}$ used in the spatial search by continuous-time QW~\cite{TSP21}.
Consider the following subsets of ${\mathcal{V}}$:
\begin{equation*}
	\nu_{\ell}=\{v\in {\mathcal{V}}:|v\cap w|=k-\ell\} \qquad (0\leqslant \ell\leqslant k),
\end{equation*}
so that $\ell$ is the length of a shortest path between $v\in \nu_\ell$ and $w$.
Note that $\nu_0=\{w\}$, and that
\begin{equation*}
	|\nu_{\ell}|=\binom{k}{\ell}\!\binom{n-k}{\ell} \qquad (0\leqslant\ell\leqslant k).
\end{equation*}
Let\footnote{Unlike in~\cite{TSP21}, we do not normalize these vectors here.}
\begin{equation*}
	\ket{\nu_{\ell}}=\sum_{v\in\nu_{\ell}} \ket{v} \qquad (0\leqslant \ell\leqslant k).
\end{equation*}
Then the subspace $\mathcal{H}_{\mathcal{V}}^{\mathrm{inv}}=\operatorname{span}\!\big\{\ket{\nu_{\ell}}:0\leqslant \ell\leqslant k\big\}$ of $\mathcal{H}_{\mathcal{V}}$ is invariant under the oracle Hamiltonian $\ket{w}\bra{w}$ and also the adjacency operator $A$ of $J(n,k)$.
In fact, by simple counting arguments we have
\begin{equation}\label{action of A}
	A\ket{\nu_{\ell}}=c_{\ell+1}\ket{\nu_{\ell+1}}+a_{\ell}\ket{\nu_{\ell}}+b_{\ell-1}\ket{\nu_{\ell-1}} \qquad (0\leqslant\ell\leqslant k),
\end{equation}
where
\begin{equation*}
	a_{\ell}=\ell(n-2\ell), \qquad b_{\ell}=(k-\ell)(n-k-\ell), \qquad c_{\ell}=\ell^2 \qquad (0\leqslant\ell\leqslant k),
\end{equation*}
and we set $\ket{\nu_{k+1}}=\ket{\nu_{-1}}:=0$ for brevity.\footnote{The numbers $a_{\ell},b_{\ell}$, and $c_{\ell}$ are known as the \emph{intersection numbers} of $J(n,k)$; see~\cite{BCN1989B,BH2012B,DKT_16}.}
Note that $d=b_0$ and that
\begin{equation}\label{sum to d}
	a_{\ell}+b_{\ell}+c_{\ell}=d \qquad (0\leqslant\ell\leqslant k).
\end{equation}

We will use the following facts about $J(n,k)$; see, e.g.,~\cite{BCN1989B,BH2012B}.
The adjacency operator $A$ has exactly $k+1$ distinct eigenvalues $\lambda_0>\lambda_1>\dots>\lambda_k$, where
\begin{equation}\label{eigenvalues}
	\lambda_{\ell}=(k-\ell)(n-k-\ell)-\ell \qquad (0\leqslant\ell\leqslant k),
\end{equation}
and the multiplicity of $\lambda_\ell$ is $\binom{n}{\ell}-\binom{n}{\ell-1}$ considering $\binom{n}{-1}=0$.
Note that $d=\lambda_0$.
For $0\leqslant \ell\leqslant k$, let $P_{\ell}$ be the projector onto the eigenspace of $A$ in $\mathcal{H}_{\mathcal{V}}$ for the eigenvalue $\lambda_{\ell}$.
Then it follows that
\begin{equation}\label{Pw}
	\big\| P_{\ell} \ket{w} \big\|^2 = \bra{w}P_{\ell} \ket{w} = \frac{\binom{n}{\ell}-\binom{n}{\ell-1}}{\binom{n}{k}}=\frac{k!(n-k)!(n-2\ell+1)}{\ell!(n-\ell+1)!}
\end{equation}
for $0\leqslant\ell\leqslant k$.
In particular, the vectors $P_{\ell} \ket{w}$ are nonzero, and hence form another orthogonal basis of the invariant subspace $\mathcal{H}_{\mathcal{V}}^{\mathrm{inv}}$.

We now turn to the discussions on the search algorithm described in Sec.~\ref{sec: search algorithm}.
Let
\begin{equation*}
	\ket{a_{\ell}} = \! \sum_{\substack{a\in {\mathcal{A}} \\ \operatorname{head}(a)\in \nu_{\ell} \\ \operatorname{tail}(a)\in \nu_{\ell}}} \!\!\! \ket{a}, \qquad \ket{b_{\ell}} = \! \sum_{\substack{a\in {\mathcal{A}} \\ \operatorname{head}(a)\in \nu_{\ell+1} \\ \operatorname{tail}(a)\in \nu_{\ell}}} \!\!\!\! \ket{a}, \qquad \ket{c_{\ell}} = \! \sum_{\substack{a\in {\mathcal{A}} \\ \operatorname{head}(a)\in \nu_{\ell-1} \\ \operatorname{tail}(a)\in \nu_{\ell}}} \!\!\!\! \ket{a},
\end{equation*}
for $0\leqslant \ell\leqslant k$.
These vectors are mutually orthogonal, but are not orthonormal.
In fact, we have
\begin{equation*}
	\big\| \ket{a_{\ell}} \big\|=\sqrt{a_\ell|\nu_{\ell}|}, \qquad \big\| \ket{b_{\ell}} \big\|=\sqrt{b_{\ell}|\nu_{\ell}|}, \qquad \big\| \ket{c_{\ell}} \big\|=\sqrt{c_\ell|\nu_{\ell}|}
\end{equation*}
for $0\leqslant\ell\leqslant k$, so that $\ket{a_0}=\ket{b_k}=\ket{c_0}=0$, and when $n=2k$ we also have $\ket{a_k}=0$.
The other vectors are always nonzero.
Moreover, we have
\begin{equation}\label{actions of U}
\begin{split}
	U\ket{a_{\ell}} &= \frac{2a_\ell}{d}\big( \ket{a_{\ell}} +\ket{b_{\ell-1}} + \ket{c_{\ell+1}}\big) -\ket{a_{\ell}}, \\
	U\ket{b_{\ell}} &= \frac{2b_\ell}{d}\big(\ket{a_{\ell}}+\ket{b_{\ell-1}}+\ket{c_{\ell+1}}\big)-\ket{c_{\ell+1}}, \\
	U\ket{c_{\ell}} &= \frac{2c_\ell}{d}\big(\ket{a_{\ell}}+\ket{b_{\ell-1}}+\ket{c_{\ell+1}}\big)-\ket{b_{\ell-1}},
\end{split}
\qquad (0\leqslant\ell\leqslant k),
\end{equation}
where we set $\ket{b_{-1}}=\ket{c_{k+1}}:=0$ for brevity.
It follows that the subspace of $\mathcal{H}_{\mathcal{A}}$ spanned by the vectors $\ket{a_{\ell}},\ket{b_{\ell}},\ket{c_{\ell}}$ $(0\leqslant \ell\leqslant k)$ is invariant under $U$.
However, we can do better.
Indeed, let $\mathcal{H}_{\mathcal{A}}^{\mathrm{inv}}$ be the $(2k+1)$-dimensional subspace of $\mathcal{H}_{\mathcal{A}}$ spanned by the vectors
\begin{equation*}
	\ket{a_{\ell}}+\ket{b_{\ell}}+\ket{c_{\ell}} \quad (0\leqslant\ell\leqslant k), \qquad \ket{b_{\ell}}-\ket{c_{\ell+1}} \quad (0\leqslant\ell<k).
\end{equation*}
These vectors are linearly independent, but do not form an orthogonal basis of $\mathcal{H}_{\mathcal{A}}^{\mathrm{inv}}$.
Note that $\mathcal{H}_{\mathcal{A}}^{\mathrm{inv}}$ has another (non-orthogonal) basis
\begin{equation*}
	\ket{a_{\ell}}+\ket{b_{\ell-1}}+\ket{c_{\ell+1}} \quad (0\leqslant\ell\leqslant k), \qquad \ket{b_{\ell}}-\ket{c_{\ell+1}} \quad (0\leqslant\ell<k).
\end{equation*}
It is now immediate to see that $\mathcal{H}_{\mathcal{A}}^{\mathrm{inv}}$ is invariant under $U$.
Moreover, the target vector $\ket{w'}$ is expressed as
\begin{equation}\label{1st expression of target vector}
	\ket{w'}=\frac{1}{\sqrt{d}}\ket{b_0},
\end{equation}
from which it follows that
\begin{gather*}
	{R}\big(\ket{a_0}+\ket{b_0}+\ket{c_0}\big)={R}\ket{b_0}=-\ket{b_0}=-\big(\ket{a_0}+\ket{b_0}+\ket{c_0}\big), \\
	{R}\big(\ket{b_0}-\ket{c_1}\big)=-\ket{b_0}-\ket{c_1}=\big(\ket{b_0}-\ket{c_1}\big)-2\big(\ket{a_0}+\ket{b_0}+\ket{c_0}\big).
\end{gather*}
It follows that $\mathcal{H}_{\mathcal{A}}^{\mathrm{inv}}$ is invariant under the oracle ${R}$, and hence under the modified evolution operator $U'=U{R}$ as well.
It is however not convenient to analyze $U'$ in terms of these bases, and we will introduce another (orthonormal) basis in the next section.

%%%%%%%%%%%%%%%%%%%
\section{An eigenbasis of the invariant subspace}\label{sec: eigenbasis}

Choose one of the arcs for every edge, and let ${\mathcal{A}}^+$ be the set of these arcs and ${\mathcal{A}}^-:={\mathcal{A}}\setminus {\mathcal{A}}^+$, so that we have
\begin{equation*}
	|\{a,\bar{a}\}\cap {\mathcal{A}}^+|=|\{a,\bar{a}\}\cap {\mathcal{A}}^-|=1 \qquad (a\in {\mathcal{A}}).
\end{equation*}
For every edge $e\in {\mathcal{E}}$ associated with an arc $a\in {\mathcal{A}}^+$, let
\begin{equation*}
	\ket{e}=\frac{1}{\sqrt{2}}\big(\ket{a}+\ket{\bar{a}}\big), \qquad \ket{\tilde{a}}=\frac{1}{\sqrt{2}}\big(\ket{a}-\ket{\bar{a}}\big).
\end{equation*}
Then these vectors form an orthonormal basis of $\mathcal{H}_{\mathcal{A}}$.
Let
\begin{equation*}
	\mathcal{H}_{\mathcal{A}}^0=\operatorname{span}\!\big\{ \ket{e} : e\in {\mathcal{E}}\big\}, \qquad \mathcal{H}_{\mathcal{A}}^1=\operatorname{span}\!\big\{ \ket{\tilde{a}} : a\in {\mathcal{A}}^+\big\},
\end{equation*}
so that $\mathcal{H}_{\mathcal{A}}=\mathcal{H}_{\mathcal{A}}^0\bigoplus\mathcal{H}_{\mathcal{A}}^1$.
We note that the subspaces $\mathcal{H}_{\mathcal{A}}^0$ and $\mathcal{H}_{\mathcal{A}}^1$ are independent of the choice of ${\mathcal{A}}^+$.
Define the linear operators $S:\mathcal{H}_{\mathcal{V}}\rightarrow\mathcal{H}_{\mathcal{A}}^0$ and $T:\mathcal{H}_{\mathcal{V}}\rightarrow\mathcal{H}_{\mathcal{A}}^1$ by
\begin{equation*}
	 S\ket{v} = \sum_{\substack{e\in\mathcal{E} \\ v\in e}} \ket{e} \qquad (v\in {\mathcal{V}}), 
\end{equation*}
and
\begin{equation*}
	 T\ket{v}= \! \sum_{\substack{a\in\mathcal{A}^+ \\ \operatorname{tail}(a)=v}} \!\! \ket{\tilde{a}} - \! \sum_{\substack{a\in\mathcal{A}^+ \\ \operatorname{head}(a)=v}} \!\!\! \ket{\tilde{a}} \qquad (v\in {\mathcal{V}}), 
\end{equation*}
respectively.
These operators are also independent of ${\mathcal{A}}^+$.
Observe that
\begin{equation}\label{L,Q arise}
	S^{\dagger}S=Q:=dI+A, \qquad T^{\dagger}T=L:=dI-A,
\end{equation}
where $L$ and $Q$ are the Laplacian operator and the signless Laplacian operator of $J(n,k)$, respectively; cf.~\cite{BH2012B}.

Recall the orthogonal basis $\big\{ \ket{\nu_{\ell}}:0\leqslant\ell\leqslant k\big\}$ of $\mathcal{H}_{\mathcal{V}}^{\mathrm{inv}}$.
We have
\begin{equation*}
\begin{split}
	S\ket{\nu_{\ell}} &= \frac{1}{\sqrt{2}}\big(2\ket{a_{\ell}}+\ket{b_{\ell}}+\ket{c_{\ell}}+\ket{b_{\ell-1}}+\ket{c_{\ell+1}}\big), \\
	T\ket{\nu_{\ell}} &= \frac{1}{\sqrt{2}}\big(\ket{b_{\ell}}-\ket{c_{\ell+1}}-\ket{b_{\ell-1}}+\ket{c_{\ell}}\big),
\end{split}
\qquad (0\leqslant \ell\leqslant k).
\end{equation*}
Hence it follows that
\begin{equation*}
	S\mathcal{H}_{\mathcal{V}}^{\mathrm{inv}}\subset \mathcal{H}_{\mathcal{A}}^0\cap\mathcal{H}_{\mathcal{A}}^{\mathrm{inv}}, \qquad T\mathcal{H}_{\mathcal{V}}^{\mathrm{inv}}\subset \mathcal{H}_{\mathcal{A}}^1\cap\mathcal{H}_{\mathcal{A}}^{\mathrm{inv}}.
\end{equation*}

Next, recall the orthogonal eigenbasis $\big\{P_{\ell}\ket{w}:0\leqslant \ell\leqslant k\big\}$ of $\mathcal{H}_{\mathcal{V}}^{\mathrm{inv}}$ for the adjacency operator $A$.
By \eqref{L,Q arise} we have
\begin{equation}\label{squared norms}
\begin{split}
	\big\| SP_{\ell}\ket{w}\big\|^2=\bra{w}P_{\ell}QP_{\ell}\ket{w}=(d+\lambda_{\ell})\big\|P_{\ell}\ket{w}\big\|^2, \\
	\big\| TP_{\ell}\ket{w}\big\|^2=\bra{w}P_{\ell}LP_{\ell}\ket{w}=(d-\lambda_{\ell})\big\|P_{\ell}\ket{w}\big\|^2,
\end{split}
\qquad (0\leqslant\ell\leqslant k).
\end{equation}
Since $\lambda_0=d$ and $\lambda_k>-d$, we have $TP_0\ket{w}=0$, and the vectors
$SP_{\ell}\ket{w}$ $(0\leqslant \ell\leqslant k)$ and $TP_{\ell}\ket{w}$ $(1\leqslant\ell\leqslant k)$ form an orthogonal basis of $\mathcal{H}_{\mathcal{A}}^{\mathrm{inv}}$.

Note that
\begin{equation}\label{S pm T}
\begin{split}
	(S+T)\ket{\nu_{\ell}} &= \sqrt{2}\,\big(\ket{a_{\ell}}+\ket{b_{\ell}}+\ket{c_{\ell}}\big), \\
	(S-T)\ket{\nu_{\ell}} &= \sqrt{2}\,\big(\ket{a_{\ell}}+\ket{b_{\ell-1}}+\ket{c_{\ell+1}}\big),
\end{split}
\qquad (0\leqslant\ell\leqslant k).
\end{equation}
Hence it follows from \eqref{action of A}, \eqref{sum to d}, and \eqref{actions of U} that
\begin{equation*}
\begin{split}
	U(S+T)\ket{\nu_{\ell}} &= (S-T)\ket{\nu_{\ell}}, \\
	 U(S-T)\ket{\nu_{\ell}} &= \frac{2}{d}(S-T)A\ket{\nu_{\ell}}-(S+T)\ket{\nu_{\ell}},
\end{split}
\qquad (0\leqslant\ell\leqslant k).
\end{equation*}
Since the $P_{\ell}\ket{w}$ are linear combinations of the $\ket{\nu_{\ell}}$, we then have
\begin{equation*}
\begin{split}
	U(S+T)P_{\ell}\ket{w} &= (S-T)P_{\ell}\ket{w}, \\
	U(S-T)P_{\ell}\ket{w} &= \frac{2\lambda_{\ell}}{d}(S-T)P_{\ell}\ket{w}-(S+T)P_{\ell}\ket{w},
\end{split}
\qquad (0\leqslant\ell\leqslant k).
\end{equation*}
Since $TP_0\ket{w}=0$, we have $USP_0\ket{w}=SP_0\ket{w}$, so that $U$ has eigenvalue $1$.
For $1\leqslant\ell\leqslant k$, the matrix representation of $U$ with respect to $(S\pm T)P_{\ell}\ket{w}$ is
\begin{equation*}
	\begin{pmatrix} 0 & -1 \\ 1 & \frac{2\lambda_{\ell}}{d} \end{pmatrix},
\end{equation*}
which has eigenvalues $\mathrm{e}^{\pm\mathrm{i}\omega_{\ell}}$, where
\begin{equation*}
	\omega_{\ell}=\arccos\!\left(\frac{\lambda_{\ell}}{d}\right).
\end{equation*}
Corresponding normalized eigenvectors are given respectively by\footnote{In fact, we can similarly construct eigenvectors of $U$ by replacing $P_{\ell}\ket{w}$ by any eigenvector of $A$ associated with eigenvalue $\lambda_{\ell}$, and then find the full spectrum of $U$ on $\mathcal{H}_{\mathcal{A}}$ as given in~\cite{HKSS14}. This method works for any regular graphs. On the other hand, the discussions in Sec.~\ref{sec: invariant subspace} are valid for any \emph{distance-regular} graphs~\cite{BCN1989B,BH2012B,DKT_16}.}
\begin{equation*}
	\ket{\omega^{\pm}_{\ell}}=\frac{\pm\mathrm{i}\sqrt{d}}{2\sqrt{d^2-\lambda_{\ell}^2}\, \big\|P_{\ell}\ket{w}\big\|}\big((\mathrm{e}^{\mp\mathrm{i}\omega_{\ell}}-1)S + (\mathrm{e}^{\mp\mathrm{i}\omega_{\ell}}+1)T\big)P_{\ell}\ket{w},
\end{equation*}
where the scaling is derived using \eqref{squared norms}.
Setting
\begin{equation*}
	\ket{\omega_0}=\frac{1}{\sqrt{2d}\, \big\|P_0\ket{w}\big\|}SP_0\ket{w},
\end{equation*}
we have the following orthonormal eigenbasis of $\mathcal{H}_{\mathcal{A}}^{\mathrm{inv}}$ for $U$:

\begin{center}
\begin{tabular}{c|c|c|c|c}
\hline\hline
Eigenvalue & $1$ & $\mathrm{e}^{\pm\mathrm{i}\omega_1}$ & $\cdots$ & $\mathrm{e}^{\pm\mathrm{i}\omega_k}$ \\
\hline
Eigenvector & $\ket{\omega_0}$ & $\ket{\omega^{\pm}_1}$ & $\cdots$ & $\ket{\omega^{\pm}_k}$ \\
\hline\hline
\end{tabular}
\end{center}

By \eqref{1st expression of target vector} and \eqref{S pm T} and since $\ket{w}=\ket{\nu_0}$, the target vector $\ket{w'}$ is written as
\begin{align}
	\ket{w'} &= \frac{1}{\sqrt{2d}}(S+T)\ket{w} \notag \\
	&= \frac{1}{\sqrt{2d}}(S+T)\sum_{\ell=0}^k P_{\ell}\ket{w} \notag \\
	&= \big\| P_0\ket{w} \big\| \ket{\omega_0} + \frac{1}{\sqrt{2}}\sum_{\ell=1}^k \big\| P_{\ell}\ket{w} \big\| \big( \ket{\omega^+_{\ell}}+\ket{\omega^-_{\ell}} \big). \label{2nd expression of target vector}
\end{align}

Observe that $P_0=\ket{s}\bra{s}$, where $\ket{s}$ denotes the uniform superposition
\begin{equation*}
	\ket{s}=\frac{1}{\sqrt{N}}\sum_{v\in {\mathcal{V}}}\ket{v}=\frac{1}{\sqrt{N}}\sum_{\ell=0}^k \ket{\nu_{\ell}},
\end{equation*}
so that we have $P_0\ket{w}=\big\| P_0\ket{w} \big\| \ket{s}$, and it follows from \eqref{S pm T} that the initial state $\ket{\psi(0)}$ is written as
\begin{equation}\label{s'=omega_0}
	\ket{\psi(0)}=\frac{1}{\sqrt{dN}}\sum_{\ell=0}^k \big(\ket{a_{\ell}}+\ket{b_{\ell}}+\ket{c_{\ell}}\big)=\frac{1}{\sqrt{2d}}(S+T)\ket{s}= \ket{\omega_0}.
\end{equation}

%%%%%%%%%%%%%%%%%%%
\section{Approximate eigenvectors of $U'$}\label{sec: eigenvectors}

Recall that $k$ is fixed.
Set
\begin{equation*}
	\epsilon:=\frac{1}{\sqrt{n}}.
\end{equation*}
From \eqref{eigenvalues} it follows that
\begin{equation*}
	\frac{\lambda_{\ell}}{d}=x_{\ell}(\epsilon):=\frac{(k-\ell)(1-(k+\ell)\epsilon^2)-\ell\epsilon^2}{k(1-k\epsilon^2)} \qquad (0\leqslant\ell\leqslant k).
\end{equation*}
Likewise, from \eqref{Pw} it follows that
\begin{equation*}
	\big\| P_{\ell}\ket{w} \big\| = p_{\ell}(\epsilon) := \epsilon^{k-\ell}\sqrt{\frac{ k! (1-(2\ell-1)\epsilon^2) }{ \ell! (1-(\ell-1)\epsilon^2)\cdots (1-(k-1)\epsilon^2) }}
\end{equation*}
for $0\leqslant \ell\leqslant k$.
We will freely use the following evaluation:
\begin{equation*}
	p_{\ell}(\epsilon)=\sqrt{\frac{k!}{\ell!}}\,\epsilon^{k-\ell}+O(\epsilon^{k-\ell+1}) \qquad (0\leqslant \ell\leqslant k).
\end{equation*}
Note that
\begin{equation*}
	\mathrm{e}^{\pm\mathrm{i}\omega_{\ell}}=e_{\ell}^{\pm}(\epsilon):=x_{\ell}(\epsilon)\pm\mathrm{i}\sqrt{1-x_{\ell}(\epsilon)^2} \qquad (1\leqslant\ell\leqslant k),
\end{equation*}
and that
\begin{equation}\label{sum to 1}
	1=\sum_{\ell=0}^k \big\| P_{\ell}\ket{w} \big\|^2= \sum_{\ell=0}^k p_{\ell}(\epsilon)^2.
\end{equation}
The matrix representation of the modified evolution operator $U'=U{R}$ with respect to the orthonormal basis $\ket{\omega_0},\ket{\omega^{\pm}_1},\dots,\ket{\omega^{\pm}_k}$ is readily obtained from the eigenvalues $1,\mathrm{e}^{\pm\mathrm{i}\omega_1},\dots,\mathrm{e}^{\pm\mathrm{i}\omega_k}$ of $U$ and \eqref{2nd expression of target vector}.

As in~\cite{TSP21}, we invoke the implicit function theorem for complex analytic functions~\cite{Krantz1992B}.
Extend for the moment the range of $\epsilon$ to complex numbers with $|\epsilon|^2<(2k-1)^{-1}$, so that the functions $p_{\ell}(\epsilon)$ and $e_{\ell}^{\pm}(\epsilon)$ are all analytic.
We fix $a\in\mathbb{C}\setminus\{0\}$, and consider the following $2k+3$ analytic functions of $2k+4$ variables $\epsilon,\xi_0,\xi_1^+,\xi_1^-,\dots,\xi_k^+,\xi_k^-,\lambda$, and $\gamma$:
\begin{align*}
	f_0(\,) &= \xi_0 - \sqrt{2} \,p_0(\epsilon)\! \left(\! \sqrt{2}\, p_0(\epsilon)\xi_0 +\sum_{\ell=1}^k p_{\ell}(\epsilon)(\xi_{\ell}^++\xi_{\ell}^-)\! \right) \!-\lambda\xi_0, \\
	f_1^{\pm}(\,) &= \xi_1^{\pm} - p_1(\epsilon)\!\left( \!\sqrt{2}\, p_0(\epsilon)\xi_0 +\sum_{\ell=1}^k p_{\ell}(\epsilon)(\xi_{\ell}^++\xi_{\ell}^-)\! \right) \!- e_1^{\mp}(\epsilon) \lambda\xi_1^{\pm}, \\[-1mm]
	& \ \, \vdots \\[-1mm]
	f_{k-1}^{\pm}(\,) &= \xi_{k-1}^{\pm} - p_{k-1}(\epsilon)\!\left( \!\sqrt{2}\, p_0(\epsilon)\xi_0 +\sum_{\ell=1}^k p_{\ell}(\epsilon)(\xi_{\ell}^++\xi_{\ell}^-)\! \right) \!- e_{k-1}^{\mp}(\epsilon) \lambda\xi_{k-1}^{\pm}, \\
	f_k^{\pm}(\,) &= \xi_k^{\pm} - p_k(\epsilon)\!\left(\! \sqrt{2}\, p_0(\epsilon)\xi_0 +\sum_{\ell=1}^k p_{\ell}(\epsilon)(\xi_{\ell}^++\xi_{\ell}^-)\! \right) \!- e_k^{\mp}(\epsilon) \lambda\xi_k^{\pm}  -(1+\mathrm{i})\gamma, \\
	f_{k+1}(\,) &= \sqrt{2}\, p_0(\epsilon)\xi_0 +\sum_{\ell=1}^k p_{\ell}(\epsilon)(\xi_{\ell}^++\xi_{\ell}^-) -1-\mathrm{i},  \\
	f_{k+2}(\,) &= \xi_0-a,
\end{align*}
where we have omitted the variables on the left-hand side for the sake of space.
The reasoning of the introduction of these functions is as follows.
A common zero of all but the last two functions means that we have
\begin{equation*}
	R\,\mathbf{x}=\lambda U^{\dagger}\mathbf{x} +\gamma \mathbf{d},
\end{equation*}
or equivalently,
\begin{equation}\label{meaning of common zero}
	U'\mathbf{x}=\lambda\mathbf{x} +\gamma U \mathbf{d},
\end{equation}
where
\begin{equation*}
	\mathbf{x}=(\xi_0,\xi_1^+,\xi_1^-,\dots,\xi_k^+,\xi_k^-)^{\mathsf{T}}, \qquad \mathbf{d}=(0,\dots,0,1+\mathrm{i},1+\mathrm{i})^{\mathsf{T}},
\end{equation*}
and $^{\mathsf{T}}$ denotes transpose.
(Here, we are identifying $U,R$, and $U'$ with their matrix representations.)
Each of the equations $f_{k+1},f_{k+2}\equiv 0$ determines a scaling of the vector $\mathbf{x}$.
While we impose two scaling constraints, we relax the constraint that $\mathbf{x}$ must be an eigenvector of $U'$ by adding the extra term $\gamma U \mathbf{d}$ involving an auxiliary parameter $\gamma$ in \eqref{meaning of common zero}.

Observe that
\begin{equation}\label{limit point}
	(\epsilon,\xi_0,\xi_1^+,\xi_1^-,\dots,\xi_k^+,\xi_k^-,\lambda,\gamma)=(0,a,0,\dots,0,1,\mathrm{i},1,0)
\end{equation}
is a zero of all the above functions.
The Jacobian matrix of these functions with respect to the $2k+3$ variables $\xi_0,\xi_1^+,\xi_1^-,\dots,\xi_k^+,\xi_k^-,\lambda,\gamma$ evaluated at the above point \eqref{limit point} is
\begin{equation*}
	\left(\begin{array}{c|ccc|cc|c|c}
		0 & 0 & \cdots & 0 & 0 & 0 & -a & 0 \\
		\hline
		0 & j_1^+ & & & 0 & 0 & 0 & 0 \\
		\vdots & & \ddots & & \vdots & \vdots & \vdots & \vdots \\
		0 & & & j_{k-1}^- & 0 & 0 & 0 & 0 \\
		\hline
		0 & 0 & \cdots & 0 & \mathrm{i} & -1 & \mathrm{i} & -1-\mathrm{i} \\
		0 & 0 & \cdots & 0 & -1 & -\mathrm{i} & 1 & -1-\mathrm{i} \\
		\hline
		0 & 0 & \cdots & 0 & 1 & 1 & 0 & 0 \\
		\hline
		1 & 0 & \cdots & 0 & 0 & 0 & 0 & 0
	\end{array}\right)
\end{equation*}
where $j_{\ell}^{\pm}:=1-e_{\ell}^{\mp}(0)$ for brevity, and the rows and columns are ordered as
\begin{equation*}
	f_0,f_1^+,f_1^-,\dots,f_k^+,f_k^-,f_{k+1},f_{k+2}, \qquad \xi_0,\xi_1^+,\xi_1^-,\dots,\xi_k^+,\xi_k^-,\lambda,\gamma,
\end{equation*}
respectively.
Note that this matrix is nonsingular since $a\ne 0$, and therefore it follows from the implicit function theorem that there exist analytic functions $\xi_0(\epsilon),\xi_1^{\pm}(\epsilon),\dots,\xi_k^{\pm}(\epsilon),\lambda(\epsilon)$, and $\gamma(\epsilon)$ of $\epsilon$ which describe the common zeros of these functions on some neighborhood of the point \eqref{limit point} as a one-parameter family of $\epsilon$.

From the equations $f_0,f_{k+1},f_{k+2}\equiv 0$ it follows that
\begin{equation}\label{xi_0}
	\xi_0(\epsilon)\equiv a,
\end{equation}
and that
\begin{equation}\label{lambda}
	\lambda(\epsilon)=1-\frac{(1+\mathrm{i})\sqrt{2}\,p_0(\epsilon)}{a} =1-\frac{(1+\mathrm{i})\sqrt{2k!}}{a}\epsilon^k+O(\epsilon^{k+1}).
\end{equation}
Likewise, from the equations $f_{\ell}^{\pm}\equiv 0$ $(1\leqslant\ell\leqslant k)$ and $f_{k+1}\equiv 0$ it follows that
\begin{equation*}
	\xi_{\ell}^{\pm}(\epsilon)=\frac{(1+\mathrm{i})p_{\ell}(\epsilon)}{1-e_{\ell}^{\mp}(\epsilon)\lambda(\epsilon)} \quad (1\leqslant \ell< k), \qquad \xi_k^{\pm}(\epsilon)=\frac{(1+\mathrm{i})(p_k(\epsilon)+\gamma(\epsilon))}{1-e_k^{\mp}(\epsilon)\lambda(\epsilon)}.
\end{equation*}
Plugging this and \eqref{xi_0} into the equation $f_{k+1}\equiv 0$, we have
\begin{equation}\label{expression for gamma}
	\gamma(\epsilon) = \frac{ 1-\frac{\sqrt{2}\,ap_0(\epsilon)}{1+\mathrm{i}}-\sum_{\ell=1}^k \!\left(\frac{p_{\ell}(\epsilon)^2}{1-e_{\ell}^-(\epsilon)\lambda(\epsilon)}+\frac{p_{\ell}(\epsilon)^2}{1-e_{\ell}^+(\epsilon)\lambda(\epsilon)}\right) }{ \frac{p_k(\epsilon)}{1-e_k^-(\epsilon)\lambda(\epsilon)}+\frac{p_k(\epsilon)}{1-e_k^+(\epsilon)\lambda(\epsilon)} }.
\end{equation}

Let $1\leqslant \ell\leqslant k$.
By \eqref{lambda}, we have
\begin{equation*}
	1-e_{\ell}^{\mp}(\epsilon)\lambda(\epsilon)=1-e_{\ell}^{\mp}(\epsilon)+\frac{(1+\mathrm{i})\sqrt{2k!}\,e_{\ell}^{\mp}(0)}{a}\epsilon^k +O(\epsilon^{k+1}),
\end{equation*}
so that
\begin{align*}
	\frac{1}{1-e_{\ell}^{\mp}(\epsilon)\lambda(\epsilon)} &= \frac{1}{1-e_{\ell}^{\mp}(\epsilon)}-\frac{(1+\mathrm{i})\sqrt{2k!}\,e_{\ell}^{\mp}(0)}{a(1-e_{\ell}^{\mp}(0))^2}\epsilon^k +O(\epsilon^{k+1}) \\
	&= \frac{1}{1-e_{\ell}^{\mp}(\epsilon)}+\frac{(1+\mathrm{i})k\sqrt{2k!}}{2a\ell}\epsilon^k +O(\epsilon^{k+1}),
\end{align*}
and hence
\begin{equation*}
	\frac{1}{1-e_{\ell}^-(\epsilon)\lambda(\epsilon)}+\frac{1}{1-e_{\ell}^+(\epsilon)\lambda(\epsilon)}=1+\frac{(1+\mathrm{i})k\sqrt{2k!}}{a\ell}\epsilon^k +O(\epsilon^{k+1}).
\end{equation*}
It follows that
\begin{align*}
	\sum_{\ell=1}^k  \!\bigg(\frac{p_{\ell}(\epsilon)^2}{1-e_{\ell}^-(\epsilon)\lambda(\epsilon)} & + \frac{p_{\ell}(\epsilon)^2}{1-e_{\ell}^+(\epsilon)\lambda(\epsilon)}\bigg) \\
	&= \sum_{\ell=1}^k p_{\ell}(\epsilon)^2+\frac{(1+\mathrm{i})\sqrt{2k!}}{a}\epsilon^k+O(\epsilon^{k+1}) \\
	&= 1+\frac{(1+\mathrm{i})\sqrt{2k!}}{a}\epsilon^k+O(\epsilon^{k+1}),
\end{align*}
where we have also used \eqref{sum to 1}.
Now \eqref{expression for gamma} becomes
\begin{align}\label{gamma}
	\gamma(\epsilon) &= \frac{ -\sqrt{2k!}\left(\frac{a}{1+\mathrm{i}}+\frac{1+\mathrm{i}}{a}\right)\!\epsilon^k+O(\epsilon^{k+1}) }{ 1+O(\epsilon) } \nonumber\\
	&= -\sqrt{2k!}\left(\frac{a}{1+\mathrm{i}}+\frac{1+\mathrm{i}}{a}\right)\!\epsilon^k+O(\epsilon^{k+1}).
\end{align}

%%%%%%%%%%%%%%%%%%%
\section{Analysis of the algorithm}\label{sec: analysis}

Recall that the results of the previous section depends on the scalar $a\in\mathbb{C}\setminus\{0\}$.
We now choose $a=\pm(1-\mathrm{i})$, so that
\begin{equation*}
	\frac{a}{1+\mathrm{i}}+\frac{1+\mathrm{i}}{a}=0.
\end{equation*}
We will use $^{\sharp}$ (resp.~$^{\flat}$) to mean functions associated with $a=1-\mathrm{i}$ (resp.~$a=\mathrm{i}-1)$), e.g., $\lambda^{\sharp},\xi_{\ell}^{\pm\flat}$.
By \eqref{lambda} and \eqref{gamma} we have
\begin{gather}
	\lambda^{\sharp}(\epsilon)=1- \mathrm{i}\sqrt{2k!}\,\epsilon^k +O(\epsilon^{k+1}), \quad \lambda^{\flat}(\epsilon)=1+ \mathrm{i}\sqrt{2k!}\,\epsilon^k +O(\epsilon^{k+1}), \label{two lambdas} \\
	\gamma^{\sharp}(\epsilon),\gamma^{\flat}(\epsilon)=O(\epsilon^{k+1}). \label{two gammas}
\end{gather}
In particular, we have
\begin{equation}\label{two lambdas are very close to 1}
	|\lambda^{\sharp}(\epsilon)|,|\lambda^{\flat}(\epsilon)|=1+O(\epsilon^{k+1}),
\end{equation}
where we recall that $\epsilon=1/\sqrt{n}$ is real and positive.

Let
\begin{equation*}
	\mathbf{x}^{\sharp}=(\xi^{\sharp}_0,\xi^{+\sharp}_1,\xi^{-\sharp}_1,\dots,\xi^{+\sharp}_k,\xi_k^{-\sharp})^{\mathsf{T}}, \quad \mathbf{x}^{\flat}=(\xi^{\flat}_0,\xi^{+\flat}_1,\xi^{-\flat}_1,\dots,\xi^{+\flat}_k,\xi_k^{-\flat})^{\mathsf{T}}.
\end{equation*}
Then we have (cf.~\eqref{limit point})
\begin{equation}\label{limits of two approximate eigenvectors}
	\mathbf{x}^{\sharp}=(1-\mathrm{i},0,\dots,0,1,\mathrm{i})^{\mathsf{T}}+O(\epsilon), \quad \mathbf{x}^{\flat}=(\mathrm{i}-1,0,\dots,0,1,\mathrm{i})^{\mathsf{T}}+O(\epsilon).
\end{equation}

Our algorithm described in Sec.~\ref{sec: search algorithm} begins in the uniform superposition (cf.~\eqref{s'=omega_0})
\begin{equation}\label{initial state in eigenbasis}
	\ket{\psi(0)} = \ket{\omega_0} = (1,0,\dots,0)^{\mathsf{T}} = \frac{1+\mathrm{i}}{4}(\mathbf{x}^{\sharp}-\mathbf{x}^{\flat})+O(\epsilon),
\end{equation}
whereas it follows from \eqref{2nd expression of target vector} that the target vector $\ket{w'}$ satisfies
\begin{equation}\label{target vector in eigenbasis}
	\ket{w'}= \frac{1}{\sqrt{2}}(0,\dots,0,1,1)^{\mathsf{T}}+O(\epsilon).
\end{equation}
Recall that the running time is
\begin{equation*}
	t_{\mathrm{run}}=\left\lfloor\frac{\pi n^{k/2}}{2\sqrt{2k!}}\right\rfloor=\frac{\pi}{2\sqrt{2k!}\,\epsilon^k}+O(1).
\end{equation*}
By \eqref{two lambdas}, we have
\begin{equation*}
	t_{\mathrm{run}} \operatorname{Log}\!\left(\frac{\lambda^{\flat}(\epsilon)}{\lambda^{\sharp}(\epsilon)}\right) = t_{\mathrm{run}} \operatorname{Log}\!\big(1+2\mathrm{i}\sqrt{2k!}\,\epsilon^k+O(\epsilon^{k+1})\big) = \mathrm{i}\pi+O(\epsilon),
\end{equation*}
so that
\begin{equation}\label{ratio of two lambdas}
	\left(\frac{\lambda^{\flat}(\epsilon)}{\lambda^{\sharp}(\epsilon)}\right)^{\!\!t_{\mathrm{run}}}=-1+O(\epsilon).
\end{equation}
On the other hand, it follows from \eqref{meaning of common zero} that
\begin{equation*}
	(U')^{t_{\mathrm{run}}}\mathbf{x}^{\sharp} = (\lambda^{\sharp}(\epsilon))^{t_{\mathrm{run}}}\mathbf{x}^{\sharp} +\gamma^{\sharp}(\epsilon)\sum_{\ell=1}^{t_{\mathrm{run}}} (\lambda^{\sharp}(\epsilon))^{\ell-1}(U')^{t_{\mathrm{run}}-\ell} U\mathbf{d}.
\end{equation*}
Pick a constant $M>0$ such that $|\lambda^{\sharp}(\epsilon)|<1+M\epsilon^{k+1}$ (cf.~\eqref{two lambdas are very close to 1}).
Then, since $U$ and $U'$ are unitary and since $(1+M\epsilon^{k+1})^{t_{\mathrm{run}}}=1+O(\epsilon)$, it follows from \eqref{two gammas} that  the norm of the second term of the right-hand side above is bounded by
\begin{align*}
	|\gamma^{\sharp}(\epsilon)| \sum_{\ell=1}^{t_{\mathrm{run}}} (1+M\epsilon^{k+1})^{\ell-1} \|\mathbf{d}\| &\leqslant |\gamma^{\sharp}(\epsilon)| t_{\mathrm{run}} (1+M\epsilon^{k+1})^{t_{\mathrm{run}}} \|\mathbf{d}\| = O(\epsilon).
\end{align*}
Hence
\begin{equation}\label{1st approximate eigenvector}
	(U')^{t_{\mathrm{run}}}\mathbf{x}^{\sharp} = (\lambda^{\sharp}(\epsilon))^{t_{\mathrm{run}}}\mathbf{x}^{\sharp} +O(\epsilon).
\end{equation}
Likewise, we have
\begin{equation}\label{2nd approximate eigenvector}
	(U')^{t_{\mathrm{run}}}\mathbf{x}^{\flat} = (\lambda^{\flat}(\epsilon))^{t_{\mathrm{run}}}\mathbf{x}^{\flat} +O(\epsilon).
\end{equation}
From \eqref{limits of two approximate eigenvectors}, \eqref{initial state in eigenbasis}, \eqref{ratio of two lambdas}, \eqref{1st approximate eigenvector}, and \eqref{2nd approximate eigenvector}, and since $U'$ is unitary, it follows that
\begin{align}
	\ket{\psi(t_{\mathrm{run}})} &= (U')^{t_{\mathrm{run}}}\ket{\psi(0)} \notag \\
	&= \frac{1+\mathrm{i}}{4}\big((\lambda^{\sharp}(\epsilon))^{t_{\mathrm{run}}}\mathbf{x}^{\sharp}-(\lambda^{\flat}(\epsilon))^{t_{\mathrm{run}}}\mathbf{x}^{\flat}\big) +O(\epsilon) \notag \\
	&= \frac{1+\mathrm{i}}{4}(\lambda^{\sharp}(\epsilon))^{t_{\mathrm{run}}}\big(\mathbf{x}^{\sharp}+\mathbf{x}^{\flat}\big) +O(\epsilon) \notag \\
	&= \frac{1+\mathrm{i}}{2}(\lambda^{\sharp}(\epsilon))^{t_{\mathrm{run}}} (0,\dots,0,1,\mathrm{i})^{\mathsf{T}} +O(\epsilon), \label{measure this state}
\end{align}
where we also note that $|\lambda^{\sharp}(\epsilon)|^{t_{\mathrm{run}}}=1+O(\epsilon)$ by \eqref{two lambdas are very close to 1}.

Let $a\in\mathcal{A}$ be such that $\operatorname{tail}(a)=w$.
Then it is easy to see that the projection of $\ket{a}$ to $\mathcal{H}_{\mathcal{A}}^{\mathrm{inv}}$ is given by $(1/\sqrt{d})\ket{w'}=(1/d)\ket{b_0}$, so that
\begin{equation*}
	\braket{a}{\psi(t_{\mathrm{run}})}=\frac{\braket{w'}{\psi(t_{\mathrm{run}})}}{\sqrt{d}}
\end{equation*}
since $\ket{\psi(t_{\mathrm{run}})}\in\mathcal{H}_{\mathcal{A}}^{\mathrm{inv}}$.
By \eqref{target vector in eigenbasis} and \eqref{measure this state}, we now compute the success probability as follows:
\begin{equation*}
	p_{\mathrm{succ}} = \!\sum_{\substack{a\in{\mathcal{A}}\\ \operatorname{tail}(a)=w}} \! \left|\braket{a}{\psi(t_{\mathrm{run}})}\right|^2 =\left| \braket{w'}{\psi(t_{\mathrm{run}})}\right|^2 =\frac{1}{2}+O(\epsilon).
\end{equation*}
This completes the analysis of our algorithm.

%%%%%%%%%%%%%%%%%%%
\section{Final remarks}\label{sec:conc}

We have analyzed the spatial search algorithm on Johnson graphs $J(n,k)$ using the coined quantum walk model. In order to achieve a fully rigorous version, we have used the arc notation for the computational basis and, in the algebraic calculations, we have invoked the implicit function theorem to approximate the relevant eigenvalues and eigenvectors of the modified evolution operator and to show that the functions are analytic and can be expanded as Taylor series. Our calculations show that for a fixed $k$ and large $n$, the optimal running time is $\pi\sqrt{N}/(2\sqrt 2)$ with asymptotic probability $1/2$, where $N$ is the number of vertices. This result is slightly different from the corresponding ones using the continuous-time quantum walk because its optimal running time is smaller by a factor of $\sqrt 2$ and the success probability by a factor of $2$. This is an indication that those quantum walk versions on the Johnson graphs are not equivalent. In any case, the quantum algorithm is quadratically faster than random walks because the classical hitting time of random walks on $J(n,k)$ is $\Omega(N)$~\cite{DKT_16,markowsky2013simple}.

Our results for $k=3$ are different from the corresponding ones reported in Ref.~\cite{XRL2019QIP}, which states that the asymptotic success probability is $1$. Using the same evolution operator, our results for the success probability for $J(15,3)$ and $J(21,3)$ as a function of the number of steps are different by a factor of 1/2 comparing to the ones described in Fig.~4 of that reference. (The caption of Fig.~4 uses the wrong parameters in $J(455,3)$ and $J(1330,3)$). It seems that the authors of Ref.~\cite{XRL2019QIP} have used a wrong definition of success probability (see the comment at the end of Sec.~\ref{sec: coined model}).

As a future work, we are interested in generalizing our method for larger sub-classes of distance-transitive graphs or distance-regular graphs and to investigate algorithms for the multimarked case.

\section*{Acknowledgements}
The work of H. Tanaka was supported by JSPS KAKENHI grant number JP20K03551.
The work of R. Portugal was supported by FAPERJ grant number CNE E-26/202.872/2018, and CNPq grant number 308923/2019-7.

%\bibliographystyle{unsrt}
%\bibliography{references}

\end{document}